\documentclass[aps,prx,superscriptaddress,reprint]{revtex4-1}
\usepackage{amsmath}
\usepackage{amssymb}
\usepackage{graphicx}
\usepackage{url}
\usepackage{hyperref}
\usepackage{color,xcolor}
\usepackage{epstopdf}
\usepackage{float}
\usepackage{ulem}
\usepackage[percent]{overpic}

\begin{document}

\title{Orbital ordering in the layered perovskite material CsVF$_4$}

\author{Ling-Fang Lin}
\affiliation{Department of Physics and Astronomy, University of Tennessee, Knoxville, Tennessee 37996, USA}
\author{Nitin Kaushal}
\affiliation{Department of Physics and Astronomy, University of Tennessee, Knoxville, Tennessee 37996, USA}
\affiliation{Materials Science and Technology Division, Oak Ridge National Laboratory, Oak Ridge, Tennessee 37831, USA}
\author{Yang Zhang}
\affiliation{Department of Physics and Astronomy, University of Tennessee, Knoxville, Tennessee 37996, USA}
\author{Adriana Moreo}
\author{Elbio Dagotto}
\affiliation{Department of Physics and Astronomy, University of Tennessee, Knoxville, Tennessee 37996, USA}
\affiliation{Materials Science and Technology Division, Oak Ridge National Laboratory, Oak Ridge, Tennessee 37831, USA}

\begin{abstract}
In strongly correlated electronic systems, several novel physical properties are induced by the orbital degree of freedom. In particular, orbital degeneracy near the Fermi level leads to spontaneous symmetry breaking, such as the nematic state in FeSe and the orbital ordering in several perovskite systems. Here, the novel layered perovskite material CsVF$_4$, with a $3d^2$ electronic configuration, was systematically studied using density functional theory and a multiorbital Hubbard model within the Hatree-Fock approximation. Our results show that CsVF$_4$ should be magnetic, with a G-type antiferromagnetic arrangement in the $ab$ plane and weak antiferromagnetic exchange along the $c$-axis, in agreement with experimental results. Driven by the Jahn-Teller distortion in the VF$_6$ octahedra that shorten the $c$-axis, the system displays an interesting electron occupancy $d_{xy}^1(d_{xz}d_{yz})^1$ corresponding to the lower nondegenerate $d_{xy}$ orbital being half-filled and the other two degenerate $d_{yz}$ and $d_{xz}$ orbitals sharing one electron per site. We show that this degeneracy is broken and a novel $d_{yz}$/$d_{xz}$ staggered orbital pattern is here predicted by both the first-principles and Hubbard model calculations. This orbital ordering is driven by the electronic instability associated with degeneracy removal to lower the energy.
\end{abstract}

\maketitle

\section{Introduction}
Perovskites have attracted considerable interest for decades because of their complex physical properties and
extensive application values. In these strongly correlated systems, several physical degrees of freedom, such as spin, charge, lattice, and orbital, are simultaneously active, either cooperating or competing. This induces exotic physical effects, such as colossal magnetoresistance in manganites~\cite{Dagotto:Prp,Dagotto:Sci}, magnetoelectricity~\cite{Sergienk:prb06,Sergienko:Prl}, electronic phase separation \cite{Dagotto:Prp,Moreo:Sci,Yunoki:Prl,miao2020direct,lin2018unexpected}, and orbital ordering~\cite{varignon2019origin,chen2013modifying,varignon2017origin,zhou2007orbital}.

Among peroskites, layered perovskite compounds are remarkable because they retain the essential features of the perovskite structure while offering higher tunability and new capabilities induced by their low-dimensional properties. The layered perovskite materials are formed by slicing perovskite slabs and inserting additional species in between layers. Among the known families of layered perovskites, there are the two major structural categories: the Ruddlesden-Popper (RP) \cite{ruddlesden1957new,ruddlesden1958compound} and the Dion-Jacobson (DJ) \cite{dion1981nouvelles,jacobson1985interlayer} families, with general formulas $A'_2A_{n-1}B_nX_{3n+1}$ and $A'A_{n-1}B_nX_{3n+1}$, respectively.

These layered perovskite systems indeed have many interesting physical properties. For the simplest $n = 1$ RP case, unconventional high temperature superconductivity was discovered in doped La$_2$CuO$_4$ \cite{anderson1987resonating} and in Sr$_2$RuO$_4$ \cite{ishida1998spin}. The so-called hybrid improper ferroelectricity was initial predicted in $n = 3$ RP layered perovskites \cite{Benedek:Prl} and later confirmed experimentally \cite{oh2015experimental,yoshida2018ferroelectric,liu2018direct}. In addition, hybrid improper ferroelectricity, as well as spin helix arrangements, were also proposed in the DJ family \cite{benedek2014origin,li2012rbbinb2o7,autieri2019persistent}.

Recently, the simplest $n$=1 RP family member $\alpha$-Sr$_2$CrO$_4$ was reported to display an interesting orbital ordering transition both on the experimental and theoretical sides, even though the precise orbital ordering configuration is still under debate \cite{ishikawa2017reversed,zhu2019n}. In this material, Cr$^{4+}$ has a $3d^2$ electron configuration and the CrO$_6$ octahedra is elongated along the $c$-axis. Surprisingly, the crystal-field splitting is reversed as compared with expectations from an elongated $c$-axis. The resulting orbital arrangement is a prerequisite for possible orbital ordering in $\alpha$-Sr$_2$CrO$_4$ \cite{ishikawa2017reversed}.

Considering the physical and structural similarities with the RP layered perovskites, analogous orbital ordering should also be obtained in DJ layered perovskites. However, to our best knowledge, there are no orbital ordering experimental results reported in DJ systems. For this reason,  finding orbital ordering physics in the DJ family from the theoretical perspective could play an important role in unifying the physical mechanisms between DJ and RP layered perovskites.

\begin{figure}
\centering
\includegraphics[width=0.48\textwidth]{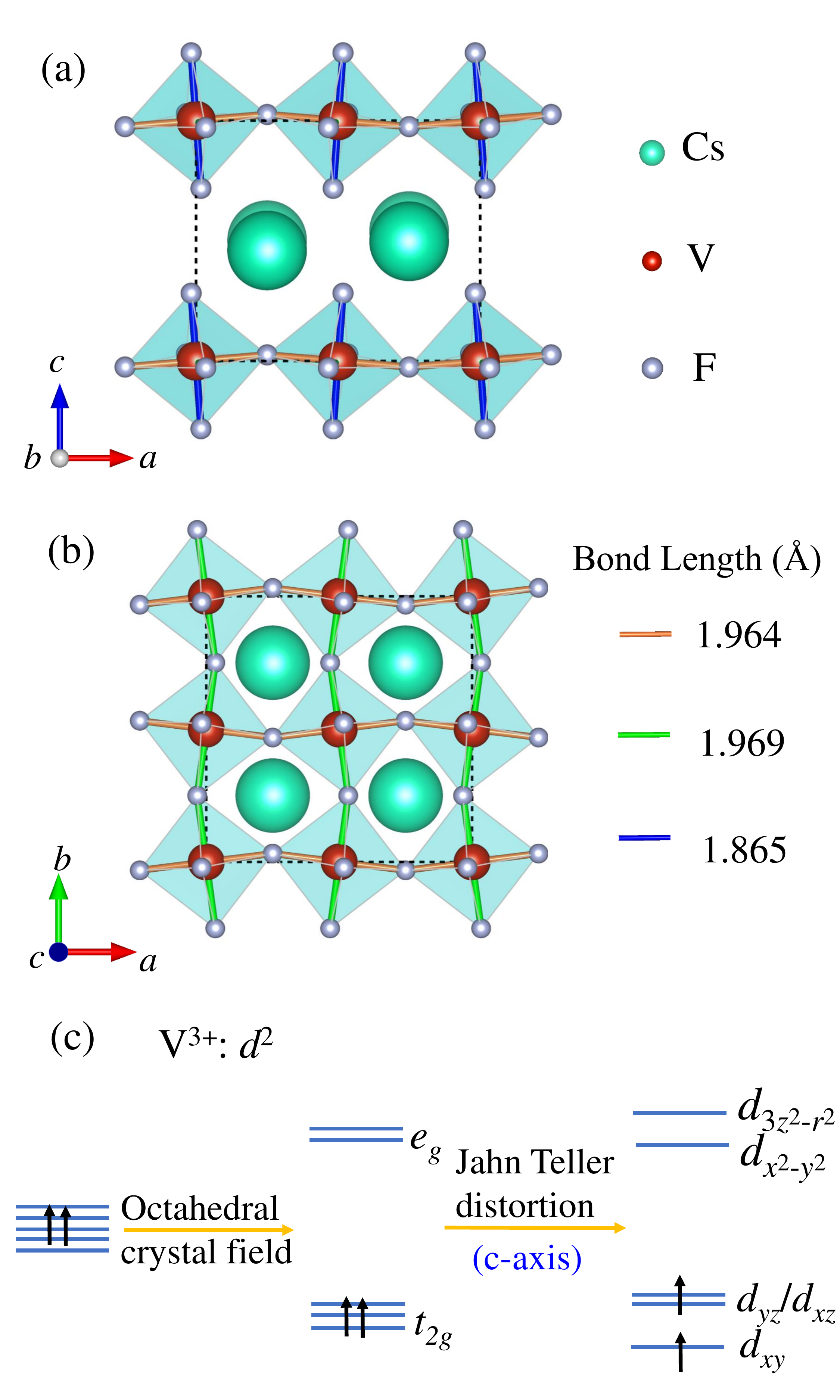}
\caption{(a) Side and (b) top views of the atomic structures for CsVF$_4$ at room temperature, respectively. Dashed rectangles indicate the unit cells. Bond lengths are indicated. (c) Schematic diagram of the expected energy-level splitting, according to the crystal-structure information available for CsVF$_4$.}
\label{structure}
\end{figure}

From the known electronic occupation configuration of $\alpha$-Sr$_2$CrO$_4$, it is reasonable to assume that finding the same $3d^2$ electronic occupation in a DJ system defines a feasible path to obtain DJ orbital-ordering physics. Hence, the layered perovskite compound CsVF$_4$ with V 3$d^2$ configuration, the simplest DJ family member corresponding to $n = 1$, captured our attention \cite{hagenmuller2012inorganic}. Fortunately, for CsVF$_4$ there is considerable and  important experimental progress. First, successive structural phase transitions have been reported for CsVF$_4$ in many investigations and the corresponding details from high to low temperature are as follows: D$_{\rm 4h}^{1}$ (phase I, space group [SG]: P4/mmm, a$^{0}$a$^{0}$c$^{0}$) $\rightarrow$ D$_{\rm 2h}^{7}$ (phase II, SG: Pman, a$_{\rm p}^{-}$a$_{\rm p}^{-}$c$^{0}$) $\rightarrow$ D$_{\rm 4h}^{7}$ (phase III, SG: P4/nmm, a$_{\rm p}^{+}$a$_{\rm p}^{+}$c$^{0}$) $\rightarrow$ D$_{\rm 2h}^{13}$ (phase IV, SG: Pmmn, a$_{\rm p}^{+}$a$_{\rm p}^{+}$c$^{+}$) $\rightarrow$ D$_{2}^{3}$ (phase V, SG: P2$_1$2$_1$2, a$_{\rm p}^{+}$b$_{\rm p}^{+}$c$^{+}$) \cite{hidaka1981polarizing,hidaka1986structural}. Second, a magnetic phase transition in CsVF$_4$ occurs at about 43~K. Due to the weak interplane coupling between the VF$_4$ layers, the magnetic structures can easily change between G- and C-antiferromagnetic (AFM) order, or a mixed state can be easily reached by applying magnetic fields \cite{hidaka1990magnetic,hidaka1996magnetic}. The most important aspect to remark is that the $3d^2$ electronic configuration provides the natural conditions for orbital ordering considering the progress reached in the study of $\alpha$-Sr$_2$CrO$_4$. Thus, it is interesting to investigate the DJ layered perovskite CsVF$_4$, especially with regard to orbital ordering, from a theoretical perspective.

In this work, the electronic and magnetic properties, as well as orbital ordering, of the simplest $n$=1 DJ layered perovskite CsVF$_4$ will be studied theoretically by using both density functional theory (DFT) and a multiorbital Hubbard model within the Hatree-Fock approximation. Due to the very weak interaction between planes in the layered structure of CsVF$_4$, this compound can be regarded as an ideal platform for quasi-two-dimensional lattice models. Our first-principles results indicate that the $t_{2g}$ orbitals of V$^{3+}$ display two one-dimensional bands originating from the $d_{xz}/d_{yz}$ orbitals and one two-dimensional band dominated by the $d_{xy}$ orbital. The G-type antiferromagnetism is found to be the magnetic ground state, with a very weak exchange coupling interaction along the $c$-axis, consistent with the expected layered structure. More interestingly, a novel staggered $d_{yz}$/$d_{xz}$ orbital-ordering pattern is here predicted, both by DFT and by the model calculations, originating in an electronic instability for the special occupancy state $(d_{xz}d_{yz})^1$. In addition, this interesting orbital-ordering pattern is sensitive to the crystal structure symmetry and could be finely adjusted by subtle distortions of the VF bonds in the $ab$ plane. Our prediction of orbital order in CsVF$_4$ also establish similarities with the orbital order discussed in manganites and ruthenates \cite{Hotta:Prb99,hotta2001prediction,csen2020properties}.

\section{Methods}
In the DFT portion of the project, first-principles calculations were performed using the revised Perdew-Burke-Ernzerhof exchange-correlation density functional (PBEsol), as implemented in the Vienna {\it ab initio} Simulation Package (VASP) code \cite{Kresse:Prb99,Blochl:Prb2,Perdew:Prl08}. The total energy convergence criterion was set to be $10^{-5}$ eV during the self-consistent calculation and the cutoff energy used for the plane-wave basis set is $550$ eV. Most calculations were carried out with the experimental crystal structure fixed, i.e. without atomistic relaxation, and the corresponding $k$-mesh employed was $4 \times 4 \times 3$. When the relaxing procedure is turned on, all lattice parameters and atomic positions were optimized to obtain the ground state structures until forces became lower than $0.01$ eV/\AA. Both non-magnetic and the spin-polarized phases were considered in our calculation. To better describe the electron correlation for the spin-polarized phase, the generalized gradient approximation plus the $U$ (GGA+$U_{\rm eff}$) approach~\cite{Dudarev:Prb} was adopted. Following previous studies addressing orbital physics in YVO$_3$ and LaVO$_3$~\cite{fang2004quantum}, the effective Hubbard coupling was fixed to the value $U_{\rm eff} = U-J = 3$ eV for simplicity. Note that, besides the correction parameter $U_{\rm eff}$, the exchange interaction is already accounted for within the spin-polarized GGA exchange-correlation potential component. Thus, it would be inappropriate to simply compare DFT results at some value of $U_{\rm eff}$ with special locations in the $J_{\rm H}/U$ and $U/W$ parameter space, as used in the model part.

From the {\it ab initio} ground-state wave function, the maximally localized Wannier functions~\cite{marzari1997maximally} within the orbital basis {$d_{xz}$, $d_{yz}$, and $d_{xy}$} for each V ion were constructed using the WANNIER90 code~\cite{mostofi2008wannier90}. Based on our well-converged {\it ab initio} calculation, the relevant hopping amplitudes and crystal-field splitting energies were extracted for the active $t_{2g}$ orbitals. Then, the ground-state phase diagram was investigated using the Hartree-Fock method based on the multi-orbital Hubbard model discussed in Sec.~\ref{model}.

\section{Lattice Properties}
According to experimental studies, there are at least five phases of the CsVF$_4$ compound, as mentioned above~\cite{hidaka1981polarizing,hidaka1986structural}. However, only the crystal structure information  measured at room temperature (phase IV) is available, corresponding to the orthorhombic symmetry (SG: Pmmn) with the lattice constants $a=7.767$, $b=7.766$, and $c=6.574$, in units of {\AA}~\cite{hidaka1986structural}. We have tried to construct the crystal structure with phase V (SG: P2$_1$2$_1$2) for CsVF$_4$ based on the information for RbFeF$_4$ at room temperature \cite{hidaka1986structuralRbFeF4} because it is is isostructural to CsVF$_4$. However, during the DFT calculation process, this initial phase V structure becomes unstable and eventually converges to the phase IV crystal structure. According to a previous study \cite{autieri2014mechanism}, high values of the Coulomb repulsion are important to stabilize the low-temperature distorted structures. Hence, our parameter $U_{\rm eff}$ was increased up to the range 4-9 eV for further testing. However, the phase V structure is still unstable during the optimization process even in this new range. Thus, almost all of our calculations were performed based on the fixed experimental structure (phase IV, SG: Pmmn) obtained at room temperature \cite{hidaka1986structural}.

As shown in Figs.~\ref{structure} (a) and (b), the CsVF$_4$ compound exhibits a single perovskite [VF$_4$]$_\infty$ layer of corner-shared VF$_6$ octahedra separated along the $c$-axis by the Cs$^{+}$ cation, forming an infinity sandwich-like structure. The tilted [VF$_6$] octahedra system is in a configuration classified as a$_{\rm p}^{+}$a$_{\rm p}^{+}$c$^{+}$ according to Glazer notation \cite{Glazer:Acb}. For each layer, the unit cell includes 4 V sites due to the rotation of the [VF$_6$] octahedra. In contrast to $\alpha$-Sr$_2$CrO$_4$, apparently each [VF$_6$] octahedron in CsVF$_4$ is shorten along the $c$ axis, while there exists only a small discrepancy between the $a$ and $b$ axes.

According to this structural information, the sketch of the expected energy-levels splitting is indicated in Fig.~\ref{structure}~(c) (the small discrepancy between $a$ and $b$ axes is ignored). Starting from the ideal cubic structure, once the [VF$_6$] octahedron is formed, the energy levels of the five $d$ orbitals splits into doubly degenerate $e_g$ orbitals and triply degenerate $t_{2g}$ orbitals. Because the [VF$_6$] octahedron is shorten along the $c$-axis, the triply degenerate $t_{2g}$ levels further split into a low-energy non-degenerate $d_{xy}$ orbital and two higher-energy doubly-degenerate $d_{xz}$ and $d_{yz}$ orbitals. V$^{3+}$ in this material has a $3d^2$ electronic configuration. Therefore, it is natural to expect that one electron always occupies the lowest $d_{xy}$ orbital while the other one is shared among the doubly degenerate $d_{xz}$ and $d_{yz}$ orbitals, which implies that the orbital degree of freedom becomes indeed active.

\section{First-principles calculations}
\begin{figure}
\centering
\includegraphics[width=0.48\textwidth]{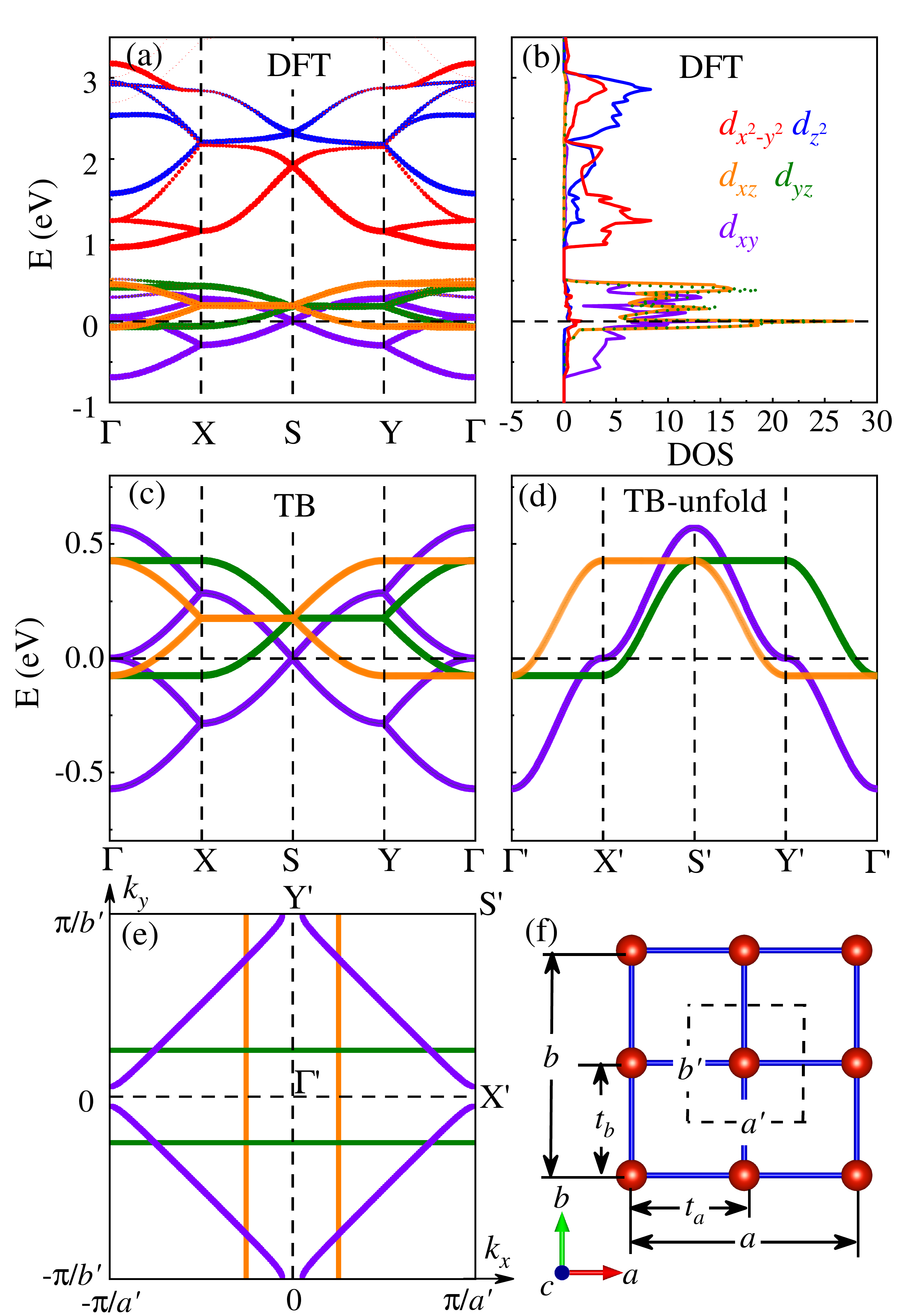}
\caption{(a) Band structure and (b) projected DOS of CsVF$_4$ from DFT calculations for non-magnetic metallic phase. (c) Tight-binding band structure. (d) Tight-binding unfolded band structure. (e) Two-dimensional FS at the $k_z$ = 0 plane in the unfolded BZ. (f) Sketch of the normal unit cell and the unfolded one. Hoppings are also indicated.}
\label{fitting}
\end{figure}

\textit{Non-magnetic Metallic Phase.}
Let us start with the hypothetical non-magnetic metallic phase of CsVF$_4$ obtained under the assumption of no spin polarization.

The band structure and projected density of states (DOS) from the DFT calculations corresponding to the V atom's $3d$ orbitals of CsVF$_4$ are shown in Figs.~\ref{fitting} (a-b). Clearly, the states near the Fermi level are mainly contributed by the $t_{2g}$ orbitals of the V ions, while the $e_{g}$ orbitals are located at a higher energy with broader bandwidth. All the three $t_{2g}$ bands crossing the Fermi energy indicates that CsVF$_4$ is a prototypical multiband system. Specifically, the energy level of the non-degenerate $d_{xy}$ orbital is lower than that of the two degenerate $d_{xz}$ and $d_{yz}$ orbitals, consistent with previous analysis of energy-level splitting, as shown in Fig.~\ref{structure} (c).

As shown in Fig.~\ref{fitting} (c), the DFT bands are fitted very well by the tight-binding (TB) bands of the three molecular orbitals obtained from the maximally localized Wannier functions. From the above TB fitting, the crystal-field levels of the $t_{2g}$ orbitals are $\Delta_{xz}$ = 0.176 eV, $\Delta_{yz}$ = 0.176 eV, and $\Delta_{xy}$ = -0.004 eV while the associated hopping amplitudes in the \{$d_{xz}$, $d_{yz}$, $d_{xy}$\} orbital basis are
\begin{equation}
\begin{split}
t_{\vec{a}} =
\begin{bmatrix}
-0.126 &     0 &      0 \\
0      &     0 &      0 \\
0      &     0 &  -0.143
\end{bmatrix},\\
t_{\vec{b}} =
\begin{bmatrix}
0      &     0 &      0 \\
0      & -0.126&      0 \\
0      &     0 &  -0.143
\end{bmatrix}.
\end{split}
\label{hoppings}
\end{equation}
Here, only the nearest-neighbor hoppings and amplitudes of hoppings larger than $0.1$ eV are considered for simplicity. To capture the degenerate properties of the $d_{xz}$ and $d_{yz}$ orbitals, reasonable modifications of the hopping parameters are adopted.

Interestingly, there are four V in the primitive unit cell in the DFT calculation. Thus, we can further simplify the current band structure by unfolding the Brillouin zone (BZ) [Fig.~\ref{fitting} (f)]. The unfolded band structure and corresponding two dimensional  Fermi surface (FS) can be found in Figs.~\ref{fitting} (d-e). According to the band structure, it is interesting that the $d_{xy}$ orbital band is broad along all high-symmetry paths in the $xy$-plane, displaying quasi-two-dimensional (2D) properties. By comparison, the $d_{xz}$ and $d_{yz}$ orbitals show almost flat bands along certain high-symmetry directions $\Gamma'$-Y$'$ (X$'$-S$'$) and $\Gamma'$-X$'$ (Y$'$-S$'$), respectively, exhibiting quasi-one-dimensional (1D) properties. Clearly, the 2D FS consists of two quasi-1D bands and one quasi-2D band. The quasi-1D bands originate from the $d_{xz}$ and $d_{yz}$ orbitals, while the quasi-2D band is dominated by the $d_{xy}$ orbital. This interesting result is quite similar to that known to occur in the chiral $p$-wave superconductor Sr$_2$RuO$_4$~\cite{kallin2012chiral}. Meanwhile, due to the 2D (1D) characteristic of the $d_{xy}$ ($d_{xz}$/$d_{yz}$) orbital, its bandwidth is broader ($4t_{d_{xy}}$) than that of the $d_{xz}/d_{yz}$ orbitals ($2t_{d_{xz}/d_{xz}}$) as shown in Figs.~\ref{fitting} (c-d).

\begin{figure}
\centering
\includegraphics[width=0.48\textwidth]{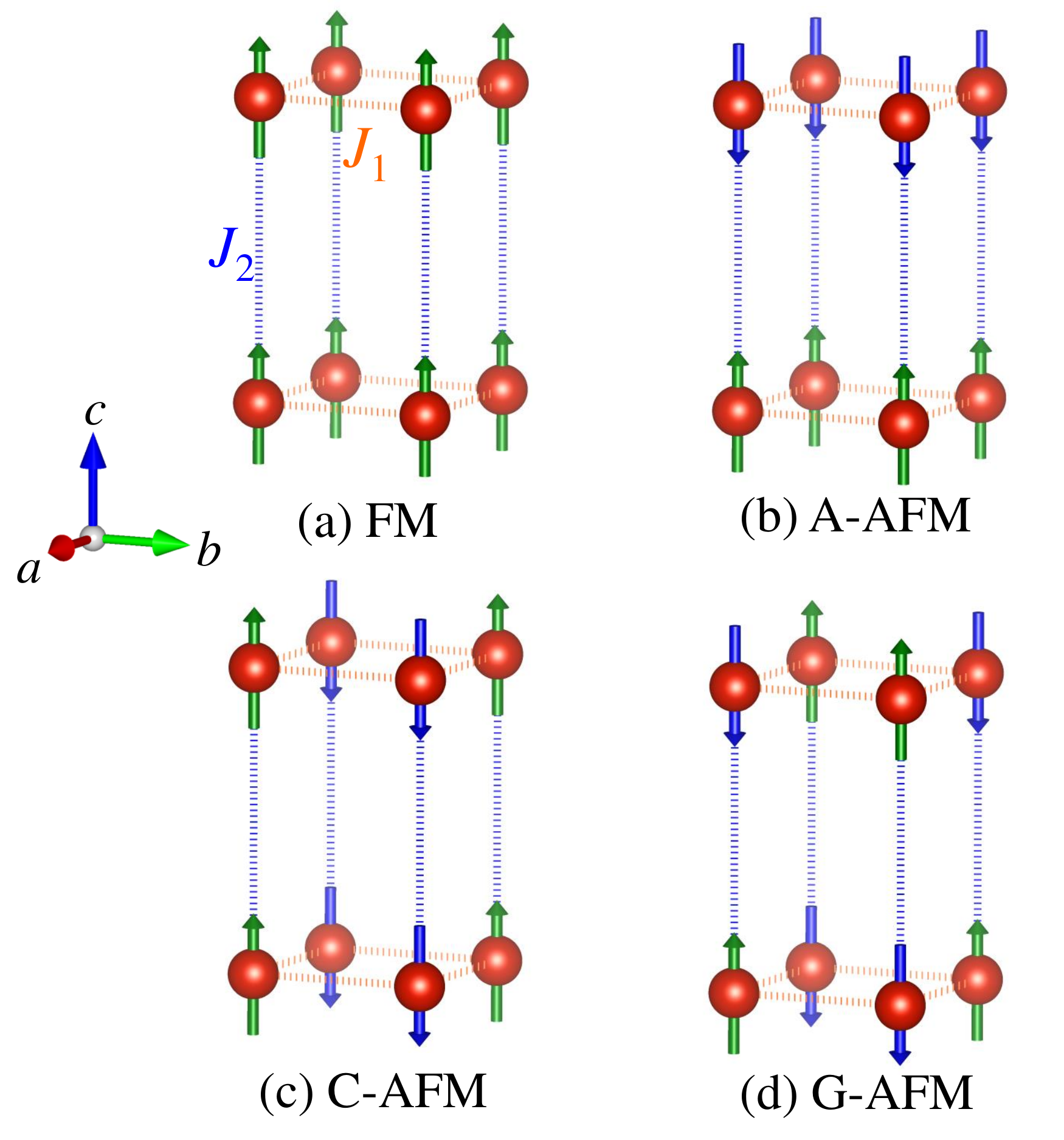}
\caption{The various magnetic configurations studied in this work. (a) FM, (b) A-AFM, (c) C-AFM, and (d) G-AFM. $J_1$ and $J_2$ are the $ab$-plane and $c$-axis exchange couplings, respectively.}
\label{magnetism}
\end{figure}

\textit{Magnetism.}
In the following DFT calculations, now the spin polarization is allowed. Four magnetic structures [i.e., ferromagnetic (FM) and A-, C-, and G-type AFM states, see Fig.~\ref{magnetism}] were calculated using the fixed atomic experimental structure discussed before. The corresponding energies are list in Table.~\ref{table1}. From this information, the exchange interactions can be estimated by mapping the calculated total energies for each magnetic state to the Heisenberg model. The nearest-neighbor exchange coupling constants can be extracted using
\begin{equation}
\begin{split}
J_1=-\frac{1}{8S^2}[E(F)-E(G)-E(C)+E(A)],\\
J_2=-\frac{1}{4S^2}[E(F)-E(G)+E(C)-E(A)],
\end{split}
\label{eq1}
\end{equation}
where $S=1$ is the magnetic moment. The extracted results are $J_1=-3.9$~meV and $J_2=-0.3$~meV indicating that both the $ab$-plane and $c$-axis favor AFM couplings. As expected, due to the layered structure of CsVF$_4$, $J_2$ is very weak and can be neglected, agreeing well with experimental investigations~\cite{hidaka1990magnetic,hidaka1996magnetic}. Since the interplane coupling is weak and we only focus on the intrinsic properties of each layer, in practice either G- or C-AFM can capture the main physics. According to our calculations below, the orbital ordering patterns are not sensitive to having G- or C-AFM magnetic order. We also tested whether the FM spin order has an effect on the orbital ordering. The results show that the orbital ordering is independent from the spin order, see the appendix Fig.~\ref{fmoo} for more details. Thus, for simplicity, the C-AFM order is considered in the following calculations, unless otherwise stated.

\begin{table}
\centering
\caption{List of energy equations and calculated energies of the four collinear spin configurations used to determine the magnetic exchange integrals. The G-AFM state is taken as the reference of energy.}
\begin{tabular*}{0.48\textwidth}{@{\extracolsep{\fill}}lccc}
\hline
\hline
Confg.& Energy equations  & Energy (meV/f.u.)\\
\hline
FM   & $E_0-2J_1S^2-J_2S^2$ & 16\\
A-AFM & $E_0-2J_1S^2+J_2S^2$ & 16\\
C-AFM & $E_0+2J_1S^2-J_2S^2$ & 1\\
G-AFM & $E_0+2J_1S^2+J_2S^2$ & 0\\
\hline
\end{tabular*}
\label{table1}
\end{table}

\begin{figure}
\centering
\includegraphics[width=0.48\textwidth]{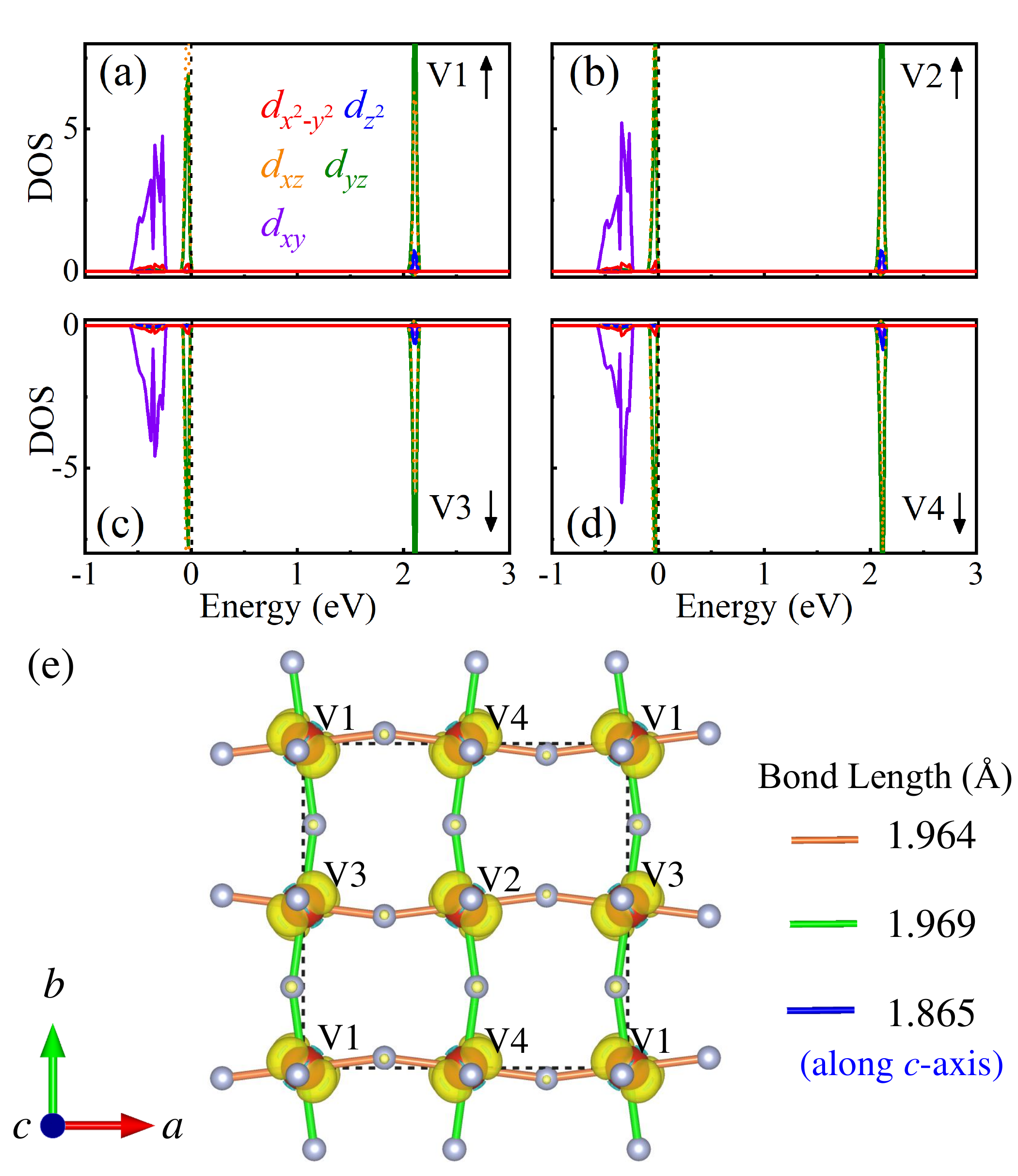}
\caption{(a)-(d) Calculated partial DOS's projected onto the five $d$ orbitals of four V ions based on the fixed experimental crystal structure with C-AFM. The vertical dash line in each panel represents the Fermi level. $\uparrow$ ($\downarrow$) represents spin up (down). (e) Charge density at the region extending from -0.2~eV to the Fermi level.}
\label{oo_exp}
\end{figure}

\textit{Orbital ordering.}
The calculated results for the projected DOS are shown in Fig.~\ref{oo_exp}, where we find that the states near the Fermi level mainly contribute from the $t_{2g}$ orbitals of the V ions while the $e_{g}$ orbitals are located at higher energies (not shown here). As expected, the $d_{xy}$ orbitals are always occupied by one electron, while the combination of degenerate $d_{xz}$ and $d_{yz}$ orbitals is occupied by another electron. An electronic instability is expected to occur when two orbitals share one electron. Therefore, linear combinations of $d_{xz}$ and $d_{yz}$ lead to two separated states, the occupied and unoccupied levels, opening a large band gap of about 2 eV. The physical reason for the band gap splitting is that the formation of the orbital ordering (OO) pattern breaks the symmetry to lower the system's energy, no matter what kind of pattern it forms. Here the large gap (2 eV) is related to the strong electronic correlation in this material, namely, the parameter $U_{\rm eff}$. A larger $U_{\rm eff}$ corresponds to a larger band gap, which is also in good agreement with the results of the Hartree-Fock model portion of this publication. The charge density for the occupied states are visually provided in Fig.~\ref{oo_exp}(e), displaying a staggered orbital ordering. Even though there are small discrepancies between the $a$- and $b$-axis lattice lengths, this anisotropy in the electronic structure and orbital ordering along the $a$ and $b$ axes can be neglected. From the symmetry point of view, when only considering the crystal symmetry (phase IV, SG: Pmmn), the V1, V2, V3, and V4 atoms are equivalent and the corresponding Wyckoff position is (0,0,0).

\begin{figure}
\centering
\includegraphics[width=0.48\textwidth]{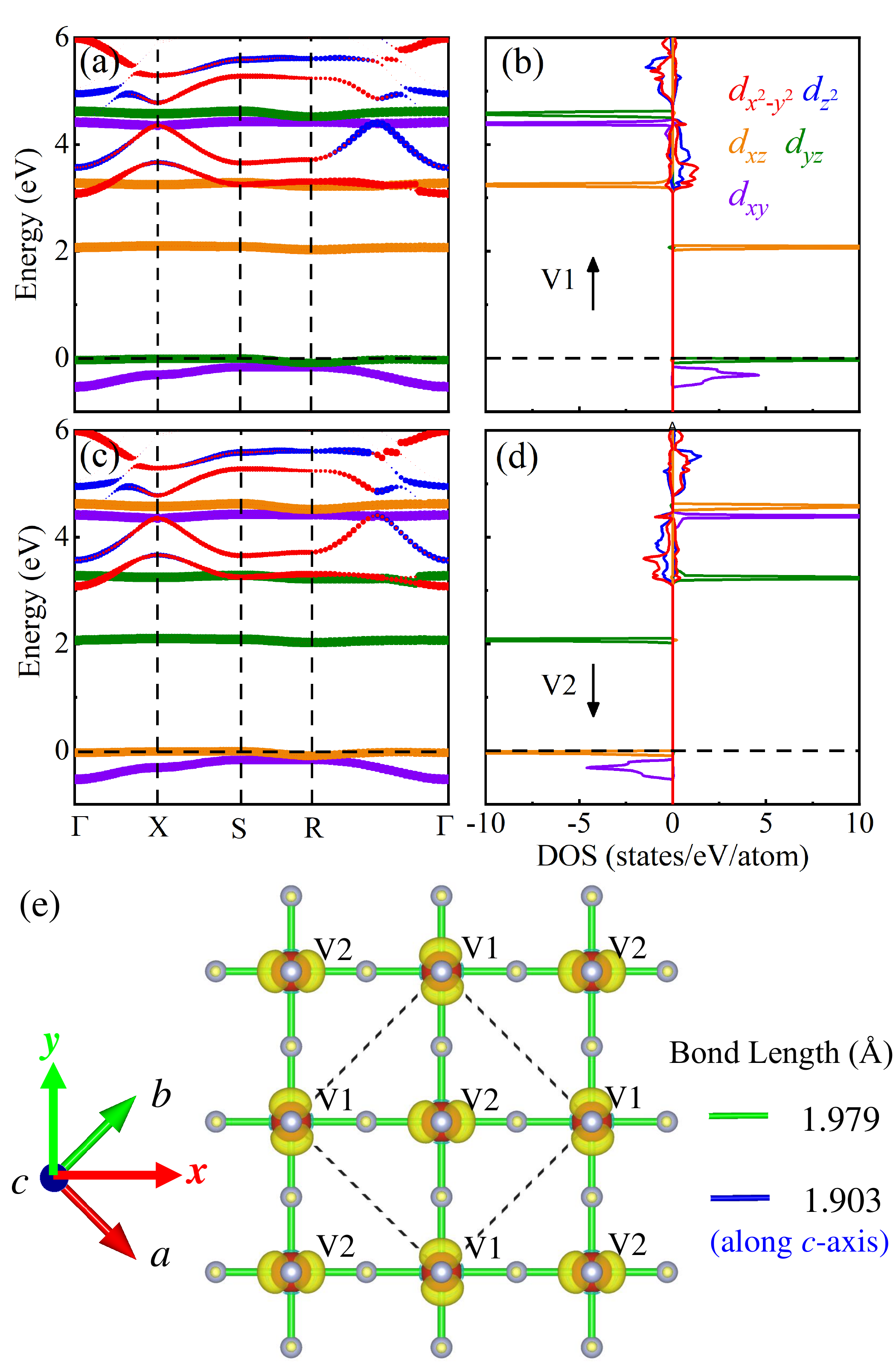}
\caption{(a)-(d) Calculated band and partial DOSs projected onto the five $d$ orbitals of four V ions based on the relaxed high-symmetry crystal structure. The horizontal dashed line in each panel represents the Fermi level. $\uparrow$ ($\downarrow$) represents spin up (down). (e) Charge density at the region extending from -0.1~eV to the Fermi level. The local axes $x$, $y$, and $z$ are defined as the [110], [$\bar{1}$10], and [001] directions of the unit cell.}
\label{oo_hss}
\end{figure}

To determine whether the electronic instability induced orbital ordering is intrinsic or not, we construct a high symmetry structure (HSS) [phase I, SG: P4/mmm, a$^{0}$a$^{0}$c$^{0})$] to remove all the distortion and rotation of the VF$_6$ octahedra in the $ab$ plane. A larger cell size ($\sqrt{2} \times \sqrt{2} \times 1$) is adopted here as compared to the minimal one (one formula unit per cell, $a=b=3.958$ and $c=6.546$ in units of {\AA} ) so as to allow for symmetry-breaking distortions. To remove interference factors, all lattice parameters and atomic coordinates are optimized for the HSS using DFT. The self-consistent calculated results are shown in Figs.~\ref{oo_hss} (a-d), with the nearly flat bands of $d_{xz}/d_{yz}$ orbitals indicating strongly localized electronic behavior. Clearly, electrons near the Fermi level occupy $d_{xz}/d_{yz}$ orbitals in a staggered manner between all nearest-neighbor V atoms in the $ab$-plane, leading to staggered orbital ordering pointing toward orthogonal directions, as shown in Fig.~\ref{oo_hss} (e). In summary, the spontaneous electronic instability unveiled here breaks degeneracies of the $d_{xz}/d_{yz}$ orbitals, resulting in staggered orbital ordering in CsVF$_4$.

\begin{figure}
\centering
\includegraphics[width=0.48\textwidth]{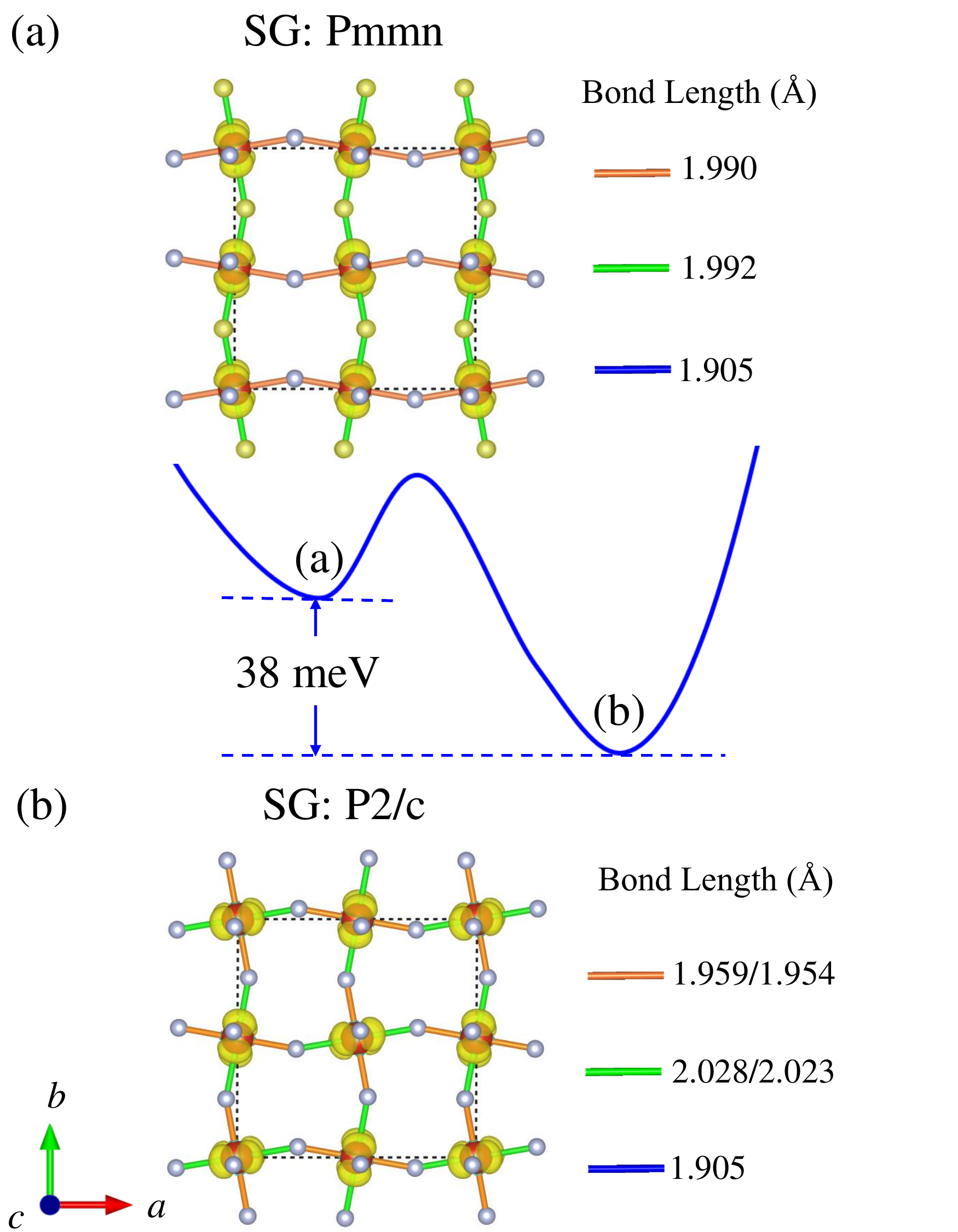}
\caption{Sketch of the (a) ferro- and (b) staggered-orbital ordering calculated with the DFT optimized structure.}
\label{oos}
\end{figure}

As mentioned before, there is a small discrepancy between the $a$- and $b$-axis if the experimental structure would be used. If we relax all lattice parameters and atomic coordinates, then the formation of ferro-orbital ordering is realized by the reinforced anisotropy [Fig.~\ref{oos} (a)]. Similarly, starting from the fixed experimental structure, if the VF bond length $ab$-plane is tuned by hand with alternating nudged amplitudes, the symmetry of the relaxed structure is lowered (SG: P2/c) as exhibited in Fig.~\ref{oos} (b). The staggered orbital ordering is reinforced and lowers the total energy by $38$ meV. According to this interesting observation, it is reasonable to speculate that the orbital ordering patterns are very sensitive to the crystal structure and controllable by fine tuning, such as via strain.

\section{\label{model}Hubbard Model}
\begin{figure}
\centering
\includegraphics[width=0.48\textwidth]{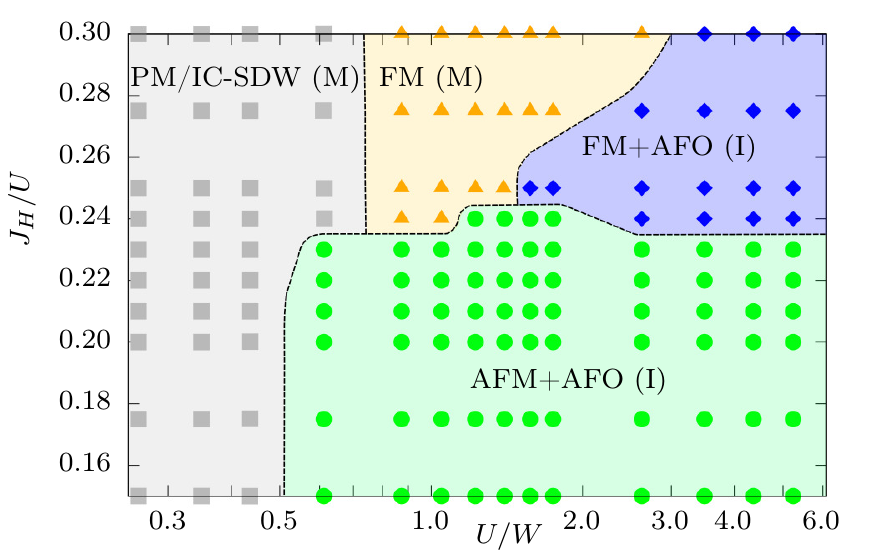}
\caption{Phase diagram of the three-orbital Hubbard model varying the Hund $J_{H}/U$ and Hubbard $U/W$ couplings, with $W$ the bandwidth. Calculations were performed for all the points shown, using a cluster size $12 \times 12$. In this cluster, the bandwidth is $W = 1.145$~eV.  The notation PM, IC-SDW, FM, AFM, AFO, M, and I stands for paramagnetic, incommensurate spin density wave, ferromagnetic, antiferromagnetic, antiferro-orbital, metallic, and insulator, respectively.
}
\label{grid}
\end{figure}

Due to the weak interaction between layers in the CsVF$_4$ compound, for simplicity only the 2D square lattice for the $ab$ plane will be considered in the electronic model. Specifically, an effective three-orbital Hubbard model for the two-dimensional square lattice will be constructed to describe the spin and orbital orderings. In all the calculations, 2 electrons per site are considered. The model studied here includes the kinetic energy and interaction energy terms $H = H_k + H_{int}$. The tight-binding kinetic component is
\begin{eqnarray}
H_k = \sum_{\substack{i\sigma\\\vec{\alpha}\gamma\gamma'}}t_{\gamma\gamma'}^{\vec{\alpha}}
(c^{\dagger}_{i\sigma\gamma}c^{\phantom\dagger}_{i+\vec{\alpha}\sigma\gamma'}+H.c.)+ \sum_{i\gamma\sigma} \Delta_{\gamma} n_{i\gamma\sigma},
\end{eqnarray}
where the first term represents the hopping of an electron from orbital $\gamma$ at site $i$ to orbital $\gamma'$ at the nearest-neighbor site $i+\vec{\alpha}$. The vector $\vec{\alpha}$ connects nearest-neighbor sites along the $\vec{a}$ and $\vec{b}$ axes, namely $\vec{\alpha}$ is the unit vector either along the $x$ or $y$ axis with length $a$ and $b$, respectively. $\gamma$ and $\gamma'$ represent the three different orbitals {$d_{xz}$, $d_{yz}$, $d_{xy}$}. $\Delta_{\gamma}$ is the crystal-field splitting of orbital $\gamma$. The actual values for the hopping matrix and crystal-field splittings are extracted from the {\it ab initio} calculations, as described in the previous section.

The electronic interaction portion of the Hamiltonian is:
\begin{eqnarray}
H_{int}= U\sum_{i\gamma}n_{i\uparrow \gamma} n_{i\downarrow \gamma} +(U'-\frac{J_H}{2})\sum_{\substack{i\\\gamma < \gamma'}} n_{i \gamma} n_{i\gamma'} \nonumber \\
-2J_H  \sum_{\substack{i\\\gamma < \gamma'}} {{\bf S}_{i,\gamma}}\cdot{{\bf S}_{i,\gamma'}}+J_H  \sum_{\substack{i\\\gamma < \gamma'}} (P^{\dagger}_{i\gamma} P_{i\gamma'}+H.c.).
\end{eqnarray}
The first term is the standard intraorbital Hubbard repulsion. The second term is the electronic repulsion between electrons at different orbitals where the standard relation $U'=U-2J_H$ is assumed. The third term represents the Hund's coupling between electrons occupying the three active $3d$ orbitals. The operator ${\bf S}_{i\gamma}$ is the total spin at site $i$ and orbital $\gamma$ defined as
\begin{eqnarray}
{\bf S}_{i\gamma}= {{1}\over{2}}\sum_{\sigma\sigma'}c^{\dagger}_{i\sigma\gamma}\sigma_{\sigma\sigma'}c^{\phantom\dagger}_{i\sigma'\gamma}.
\end{eqnarray}
The fourth term is the pair hopping between different orbitals at the same site $i$, where $P_{i\gamma}$=$c_{i \downarrow \gamma} c_{i \uparrow \gamma}$.

The unrestricted real-space Hartree-Fock method is applied to solve numerically the Hamiltonian we constructed \cite{Luo:Prb}. We performed a Hartree-Fock decomposition on all the quartic fermionic terms in the interaction, leading to the single-particle density matrix elements $\langle c_{i\sigma\gamma}^{\dagger}c_{i{\sigma}'{\gamma}'} \rangle$, as the mean-field parameters. Then, self consistency in those mean-field parameters was achieved using the modified Broyden's method \cite{johnson1988modified}. The chemical potential $\mu$ was tuned to target the required electronic density. Up to 15 random configurations of order parameters were used to start the iterative process to gain convergence at every point, and the converged states with the lowest energy were chosen as the result. We calculated the local electronic density, density of states, local spin moment $\langle  {\bf{S}}^2\rangle$, spin structure factor $S(\bf{q})$, and orbital structure factor $\tau({\bf{q}})$ to identify the phases. We used the following definitions:
\begin{eqnarray}
\langle {\bf{S}}^2\rangle&=&\frac{1}{L_{x}L_{y}}\sum_{{{i}}}\langle {\bf{S}}_{{i}}^2 \rangle, \nonumber \\
S({\bf{q}})&=&\frac{1}{(L_{x}L_{y})^2}\sum_{{i,j}} \langle {{\bf{S}}_{i}}\cdot {{\bf{S}}_{j}} \rangle e^{i {\bf{q}}\cdot ({\bf{r}}_{i}-{\bf{r}}_{j})}, \nonumber \\
\tau({\bf{q}})&=&\frac{1}{(L_{x}L_{y})^2}\sum_{{i,j}} \langle {{{\tau}}_{i}} {{{\tau}}_{j}} \rangle e^{i {\bf{q}}\cdot ({\bf{r}}_{i}-{\bf{r}}_{j})},
\end{eqnarray}
where ${\tau_{i}}=(n_{i,xz} - n_{i,yz})/2$.

\begin{figure}
\centering
\includegraphics[width=0.48\textwidth]{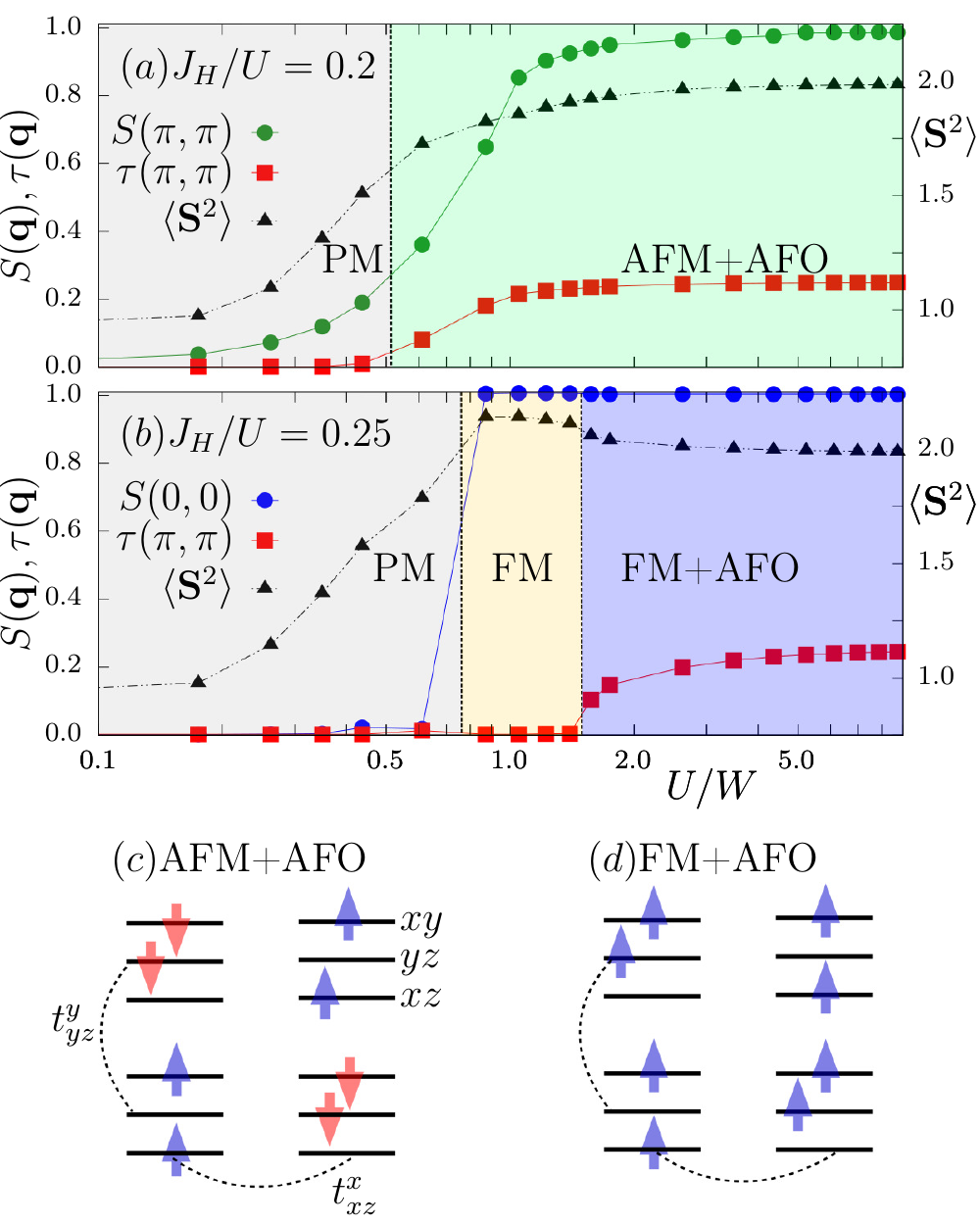}
\caption{Panels (a,b) show the spin structure factor $S(\mathbf{q})$, orbital structure factor $\tau(\mathbf{q})$, and averaged local spin moment $\langle \mathbf{S}^2 \rangle$, at $J_{H}/U=0.2$ and $0.25$. Panels (c) and (d) show the pictorial represention of the $\textrm{AFM+AFO}$ and $\textrm{FM+AFO}$ states, respectively.  }
\label{HF}
\end{figure}

Figure ~\ref{grid} shows the full phase diagram of the three-orbital Hubbard model varying $J_{H}/U$ from $0.15$ to $0.30$ and $U/W$ from $0$ to $6$. For small $U/W \le 0.6$, the system mostly shows paramagnetism and the presence of incommensurate spin density wave order near the phase boundaries. Interestingly, for $J_{H} \le 0.24$, and for most of the region of the phase diagram, we found large peaks at momentum ${\bf{q}}=(\pi,\pi)$ in $S(\bf{q})$ and $\tau(\bf{q})$ suggesting antiferromagnetic spin ordering accompanied by antiferro-orbital ordering (namely a combined state AFM+AFO). In Fig.~\ref{HF}(a), we fix $J_{H}/U=0.2$ and show the evolution of $S(\pi,\pi)$, $\tau(\pi,\pi)$, and $\langle {\bf{S}}^{2}\rangle$ with $U/W$, to illustrate that starting from intermediate Hubbard repulsion the system is in an AFM+AFO state with robust local spin moments and as we increase $U/W$ further, $\langle {\bf{S}}^2 \rangle$ saturates to $2.0$, corresponding to spin 1 as expected.

As shown in the phase diagram, for $J_{H}/U \ge 0.24$ the system mainly presents ferromagnetic ordering (FM). In Fig.~\ref{HF}(b), the evolution with $U/W$ of $S(0,0)$, $\tau(\pi,\pi)$, and $\langle {\bf{S}}^{2}\rangle$  are shown. For intermediate values of $U/W$, a FM-metallic region was found, while for large $U/W$, a FM-insulator accompanied with antiferro-orbital ordering (FM+AFO) region is present. Once again, in the FM+AFO region the spin-moment squared is saturated to value 2.0, whereas in the FM-metal region $\langle {\bf{S}}^{2} \rangle$ is slightly greater than 2.0, because of considerable contributions arising from the states with three electrons at the same site.

According to the full phase diagram, the AFM+AFO state is predicted to be the most relevant phase for the real material, because the majority of the phase diagram comprises of this state. Moreover, this is in good agreement with the DFT results. The single-particle density of states is also calculated and the system is found to be insulating in the AFM+AFO state. The average orbital-resolved local density calculations show that the $xy$ orbital is exactly half-filled (i.e $\langle n_{xy} \rangle=1$), whereas $\langle n_{xz}\rangle=\langle n_{yz} \rangle=0.5$. In Fig.~\ref{HF}(c), a pictorial representation of the AFM+AFO state is  displayed. The antiferromagnetic spin order is driven by the half-filled $xy$ orbital with the largest hopping amplitude. These $xy$ spins being parallel to the spins on orbitals $xz$/$yz$ because of the robust Hund's coupling, create the spin 1 local moment. The staggered orbital ordering among the $xz$/$yz$ orbitals is energetically preferred to ease the movement of the electrons (i.e. decrease in kinetic energy). Note that if the Hund's coupling is increased beyond $0.24$, the FM state is stabilized because now the Hund's term play the dominant role in the energy of the intermediate state via hopping of electrons [see Fig.~\ref{HF}(d)], as in the double-exchange mechanism.

\section{Discussion}

The electron-electron interaction and electron-phonon coupling are the two major possible mechanisms to cause the orbital ordering discussed here. But which one is the primary cause? Both in the DFT and model portions of the manuscript, we constructed the high symmetry structure, removing all the distortions and rotations in the $ab$ plane to analyze the role of the electron-phonon coupling in the system. Interestingly, the results show that the staggered orbital ordering is still robust. In other words, the AFO pattern dominates even when the electron-phonon coupling is not included in the model we studied. These results indicate that electron-electron interaction is the intrinsic driving mechanism in CsVF$_4$.

However, we cannot establish if spin or orbital are the main drivers of the symmetries broken. Both are entangled. Only a calculation including finite temperature can find out which of the two critical temperatures, i.e. $T_{\rm orbital}$ or $T_{\rm spin}$, occurs first upon cooling. Then that would establish which one is the ``driver" and which one the ``passenger". But this calculation is very difficult, particularly within DFT, and it is postponed to future work.

If the electron-phonon coupling would be included, how does this coupling affect the orbital ordering? Does this new coupling establish clearly whether spin or orbital dominate? These questions deserve further work. Typically, orbital and lattice work together to induce orbital order and probably with phonons included, the orbital would be the main driver over spin.

\section{Conclusions}
In this work, first-principles DFT and Hubbard model calculations for CsVF$_4$ were performed. Due to the layered structure of CsVF$_4$, the coupling between interplanes is very weak and can be neglected. For this reason, the CsVF$_4$ compound provides an ideal platform to study quasi two-dimensional lattice models. Our theoretical results indicate that the $t_{2g}$ obitals of V$^{3+}$ dominate and display two quasi one-dimensional bands originating from the $d_{xz}/d_{yz}$ orbitals and one two-dimensional band dominated by the $d_{xy}$ orbital. Furthermore, the G-type antiferromagnetism is found to be the dominant magnetic ground state, in agreement with previous experimental results. More interestingly, a novel staggered $d_{yz}$/$d_{xz}$ orbital ordering pattern is here predicted, driven by an electronic instability for the special electron occupancy state $(d_{xz}/d_{yz})^1$. In addition, this  orbital ordering pattern is sensitive to the crystal structure symmetry and could be finely adjusted by subtle distortions of the VF bond in the $ab$ plane.

\section{Acknowledgments}
This project was supported by the U.S. Department of Energy (DOE), Office of Science, Basic Energy Sciences (BES), Materials Science and Engineering Division.

\section{Appendix}
\begin{figure}
\centering
\includegraphics[width=0.48\textwidth]{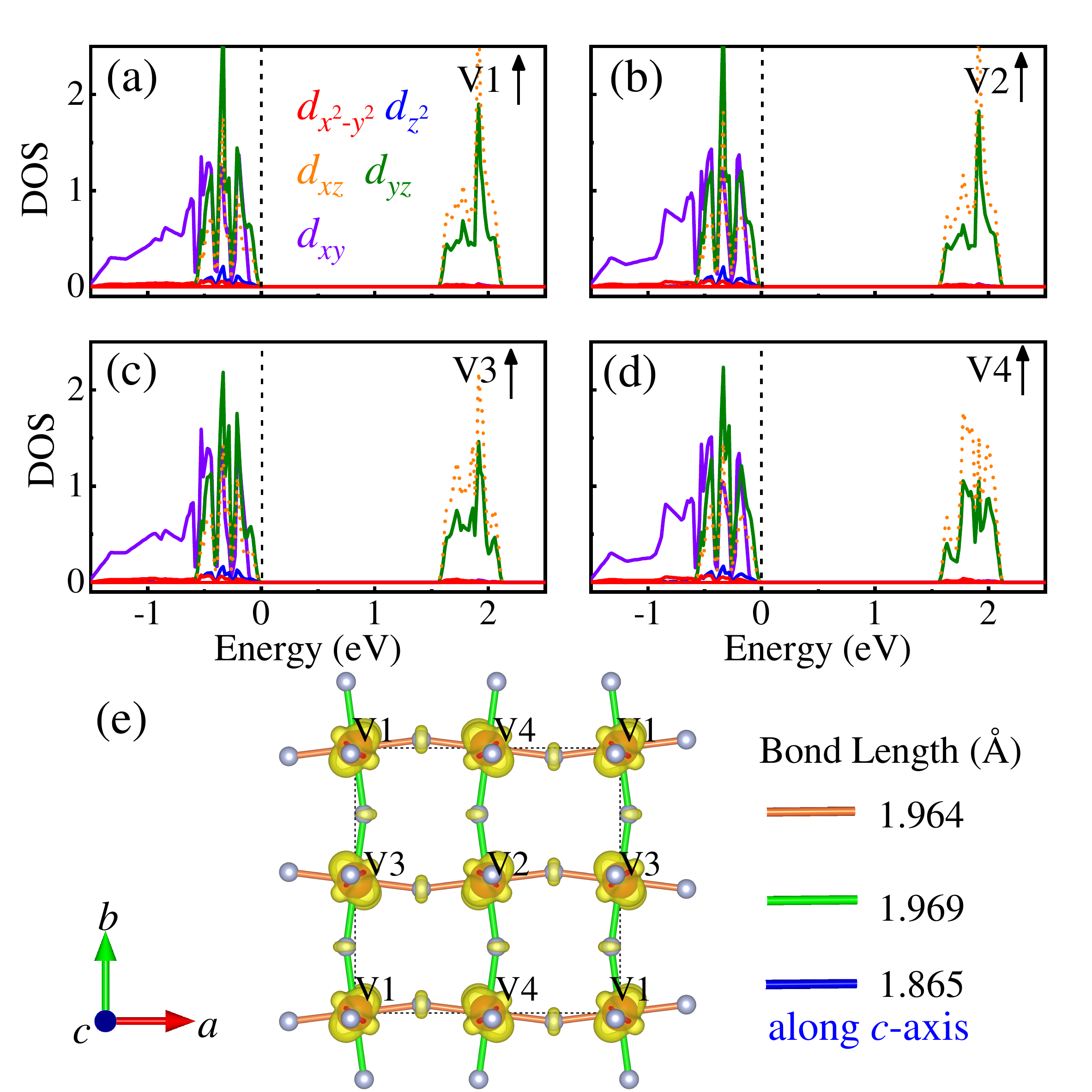}
\caption{(a)-(d) Calculated partial DOS's projected onto the five $d$ orbitals of four V ions based on the fixed experimental crystal structure with FM order. The vertical dash line in each panel represents the Fermi level. $\uparrow$ ($\downarrow$) represents spin up (down). (e) Charge density at the region extending from -0.4~eV to the Fermi level.}
\label{fmoo}
\end{figure}

As shown in Figs.~\ref{fmoo} (a-d), if the magnetism is fixed to be FM, the charge density from  the $d_{xy}$ and $d_{xz}/d_{yz}$ orbitals is not too different from the C-AFM case. Even though the final charge density shown in Fig.~\ref{fmoo} (e) is mixed with some $d_{xy}$ orbital at the region in [-0.4, 0] eV, it is clear that the $d_{xz}/d_{yz}$ orbitals are showing the same pattern as for C-AFM. In other words, the orbital ordering appears independent from the spin order. This result is also consistent with our model calculations, where both the AFM+AFO and FM+AFO phases are shown in the phase diagram to be stable at different regions.

\begin{figure}
\centering
\includegraphics[width=0.48\textwidth]{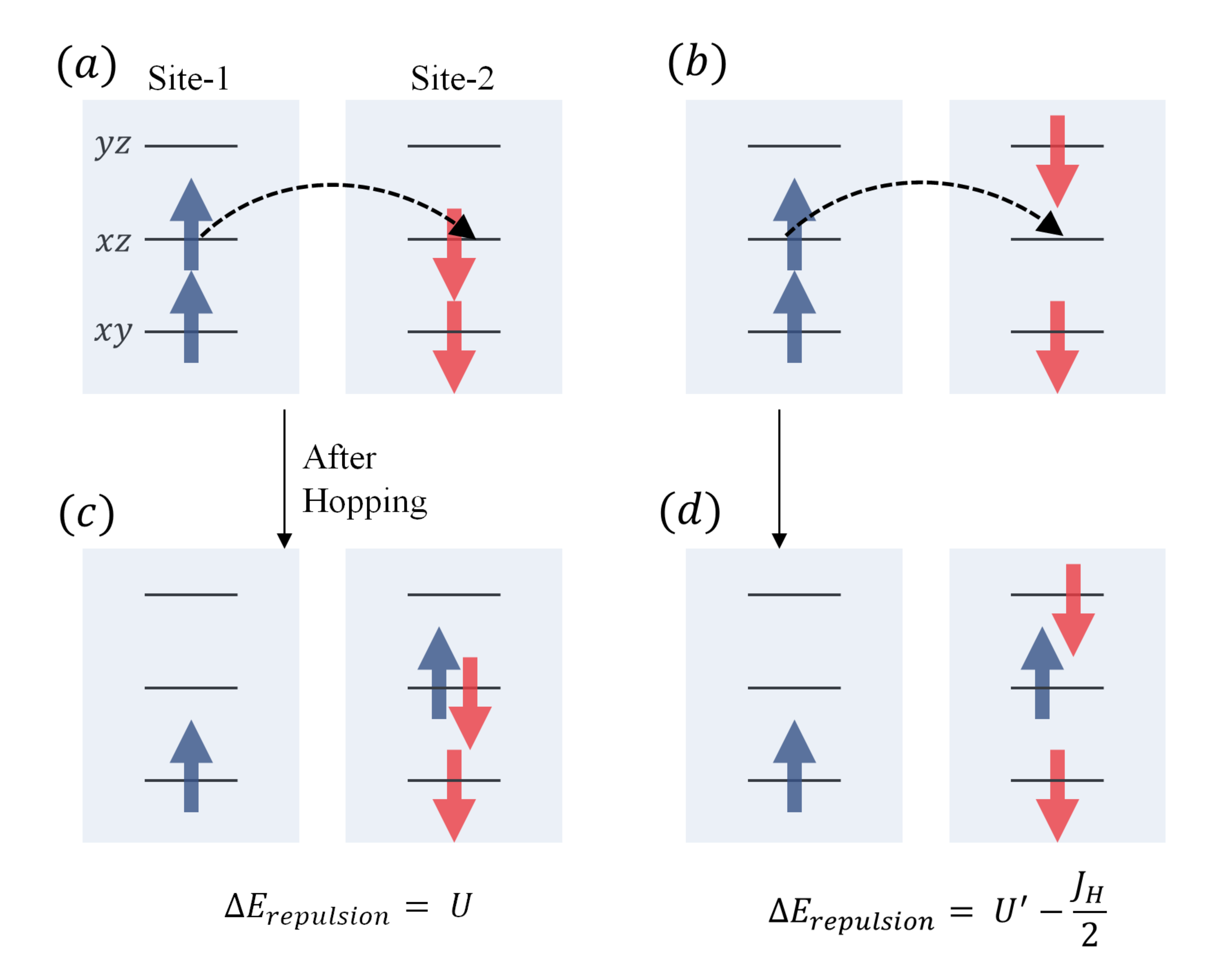}
\caption{Pictorial understanding of the $\textrm{AFM+AFO}$ states, as discussed in the text.}
\label{afo_afm}
\end{figure}
To understand why a $U/W$ robust is needed for the orbital ordering, we can start with the atomic limit. Large $U$ and robust Hund's coupling (for example $J_{\rm H}/U=0.2$) will prefer that electrons are present in different orbitals but with the same spin. The lower energy of the $d_{xy}$ orbital induces one electron to be located in the $d_{xy}$ orbital. Now, if we turn on the kinetic energy term (for simplicity we are using only two sites here) the effective superexchange between the half-filled $xy$-sites lead to antiferromagnetic ordering, while electrons in the $d_{xz}$/$d_{yz}$ orbitals just follow the same spin ordering because of the robust Hund's coupling. Now assuming this antiferromagnetic state, one of the two states in Figs.~\ref{afo_afm}(a,b) is possible. The cost of hopping as in the arrow is smaller for state in (b) because there is no double occupancy (the change in Hund's coupling energy is ignored for simplicity, as it will be same in both cases). The above discussion explains why the AFO+AFM state is favored. A similar argument can be used for the 2d lattice with the actual hopping terms used in the model; we have indeed compared the energies of the different Ansatz states using our Hartree-Fock code to understand why AFO state is preferred.

\begin{figure}
\centering
\includegraphics[width=0.48\textwidth]{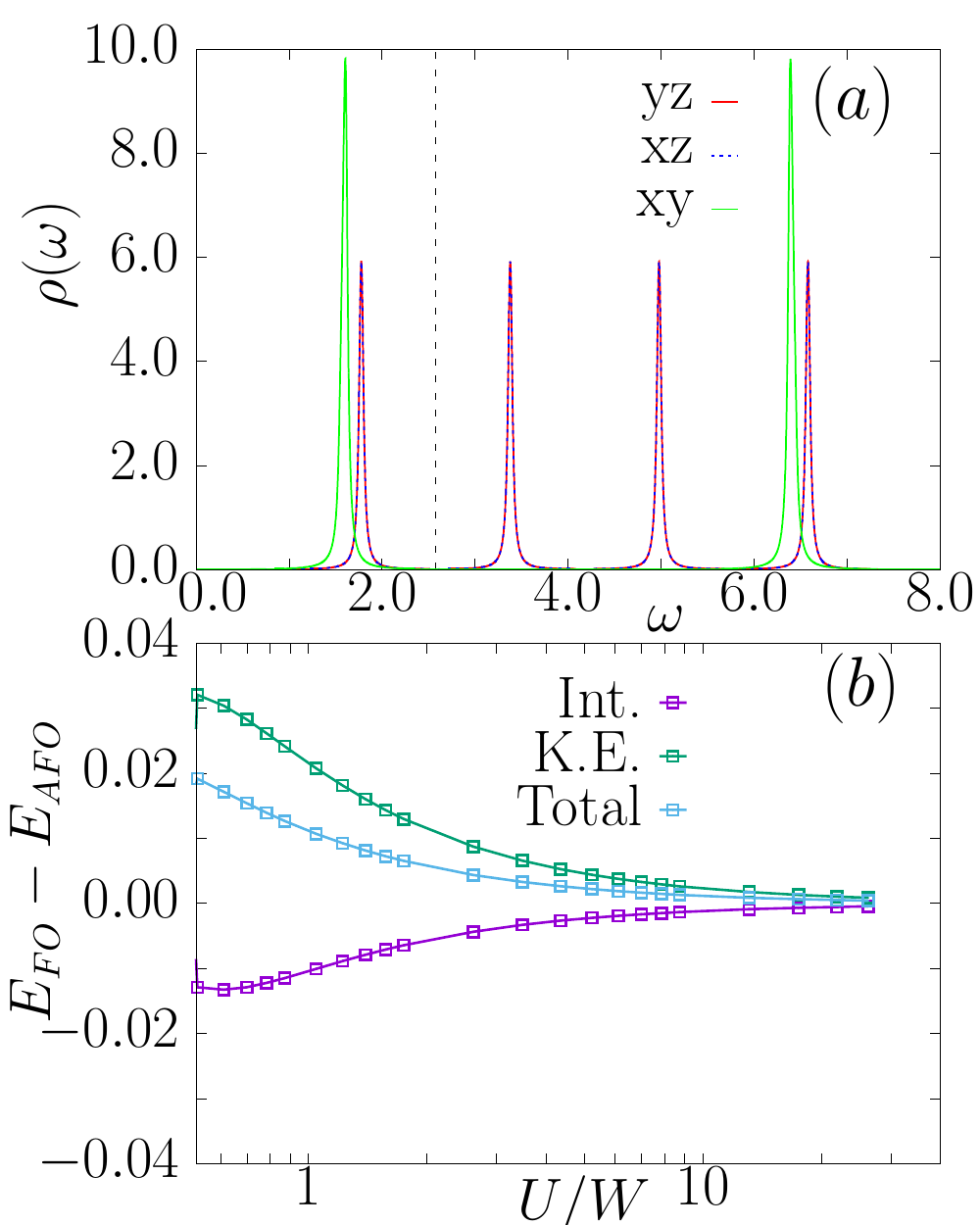}
\caption{(a) DOS for a $16 \times 16$ cluster in the AFM+AFO state with $U/W=4.0$. (b) The energies of the FO and AFO Ansatz states, both with AFM ordering.}
\label{hf_dos_en}
\end{figure}
Figure~\ref{hf_dos_en}(a) shows the DOS for a $16 \times 16$ cluster in the AFM+AFO state with $U/W=4.0$. We found a gap of nearly 1.6~eV, and we checked that the gap increases as we increase $U$. Thus, clearly correlations effects are responsible for the physics we found. We have calculated the energies of the FO and AFO Ansatz states, both with AFM ordering, as shown in panel (b). Please notice that the total energy of the FO state is higher than the AFO state, and the main reason originates in the higher kinetic energy in the FO state.

\begin{figure}
\centering
\includegraphics[width=0.48\textwidth]{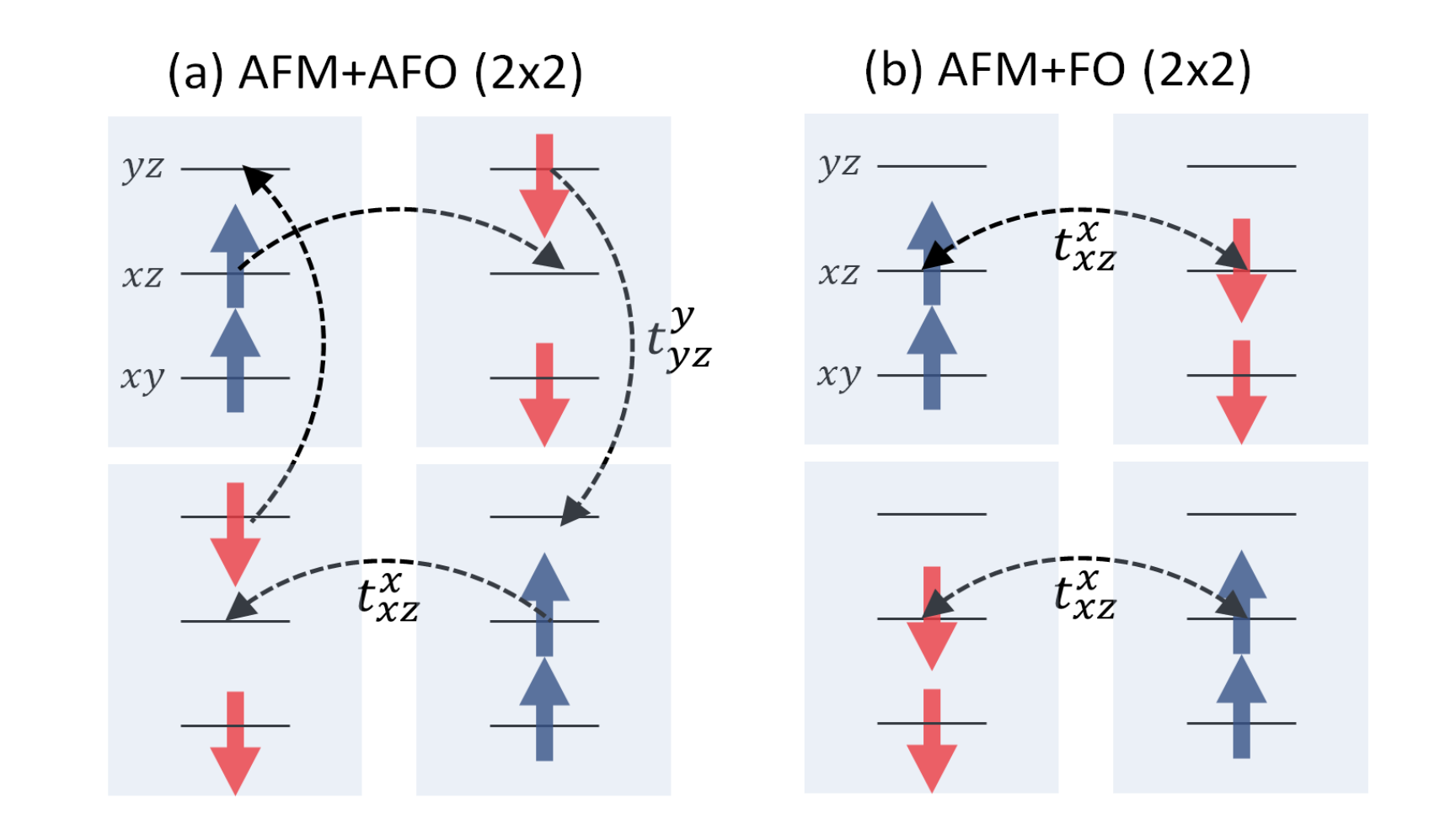}
\caption{The pictorial understanding of the AFO state using a 2$\times$2 cluster.}
\label{cartoon}
\end{figure}
Why the AFO state has lower kinetic energy i.e. why electrons move relatively easier in AFO than in FO? Intuitive understanding can be gained by the cartoon shown in Fig.~\ref{cartoon}. We can focus only on the $xz/yz$ orbitals, because the $xy$ orbital behaves similarly in both states. In the AFO state electrons can hop in both directions, whereas in the FO state the drawn-above electron hopping is restricted only to the $x$ direction because the $y$-direction hopping of the $xz$ orbital is zero. The picture described above shows that the hoppings for CsVF$_4$ further stabilizes the AFO state for a large range of $U$, in addition to the fact that in the large $U$ limit the AFO exchange is larger than the FO exchange, as discussed above in Fig.~\ref{afo_afm}.

\bibliographystyle{apsrev4-1}
\bibliography{ref3}
\end{document}